\pdfoutput=1
\documentclass{article}
\usepackage{arxiv}

\usepackage{graphicx}
\usepackage[intlimits]{amsmath}
\usepackage{amssymb}
\usepackage{exscale}
\usepackage{hyperref}
\usepackage{times}
\usepackage{subfigure}
\usepackage{subfiles}
\usepackage{wrapfig}
\usepackage{floatrow}

\title{Processing of nanostructured bulk Fe-Cr alloys by severe plastic deformation}

\author{
  Lukas Weissitsch \\
  Erich Schmid Institute of Materials Science, Austrian Academy of Sciences\\
  Jahnstra{\ss}e 12, 8700 Leoben, Austria \\
  \texttt{lukas.weissitsch@oeaw.ac.at}
\And
     Martin St\"uckler \\
  Erich Schmid Institute of Materials Science, Austrian Academy of Sciences\\
  Jahnstra{\ss}e 12, 8700 Leoben, Austria \\
\And
   Stefan Wurster \\
     Erich Schmid Institute of Materials Science, Austrian Academy of Sciences\\
  Jahnstra{\ss}e 12, 8700 Leoben, Austria\\ 
\And
  Reinhard Pippan\\
    Erich Schmid Institute of Materials Science, Austrian Academy of Sciences\\
  Jahnstra{\ss}e 12, 8700 Leoben, Austria 
\And
  Andrea Bachmaier\\
  Erich Schmid Institute of Materials Science, Austrian Academy of Sciences\\
  Jahnstra{\ss}e 12, 8700 Leoben, Austria 
}

\begin{document}
\maketitle

\begin{abstract}
The processing of binary alloys consisting of ferromagnetic Fe and antiferromagnetic Cr by severe plastic deformation (SPD) with different chemical compositions has been investigated. Although the phase diagram exhibits a large gap in the thermodynamical equilibrium at lower temperatures, it is shown that techniques based on SPD help to overcome common processing limits. 
Different processing routes including initial ball milling (BM) and arc melting (AM) and a concatenation with annealing treatments prior to high-pressure torsion (HPT) deformation are compared in this work. Investigation of the deformed microstructures by electron microscopy and synchrotron X-ray diffraction reveal homogeneous, nanocrystalline microstructures for HPT deformed AM alloys. 
HPT deformation of powder blends and BM powders leads to an exorbitant increase in hardness or an unusual fast formation of a $ \sigma$-phase and therefore impede successful processing.
\end{abstract}


\section{Introduction}

Despite the fact that binary Fe-Cr alloys are intensively studied in the past, several characteristics of this system are still not fully understood. 
Good mechanical properties in combination with great corrosion resistance \cite{mortimer_oxidation} are known, but also the prominent 475°C embrittlement \cite{grobner_embrittlement} and the $\sigma$-phase formation in a narrow compositional region \cite{dubiel_sigma,joubert_simga} in a more negative manner. Whilst details of the phase diagram are currently discussed and adapted \cite{jacob_phase_1, xiong_phase_2}, a description of spin glass formation \cite{burke_spinglass} and a strong swelling resistance under irradiative environments can be found in literature \cite{gelles_radiation_1, filippova_radiation_2}, increasing the versatility of this system.

At lower temperatures, the elemental Fe and Cr exhibit the same bcc crystal structure, but their magnetic properties differ significantly: Fe is ferromagnetic, Cr conversely is antiferromagnetic. Since the fabrication possibilities reached microstructural sizes in the nanometer regime, a break-trough in processing thin films of Fe-Cr was made. These materials are well known due to a discovery named the giant magnetoresistance \cite{baibich_gmr, binasch_gmr}. Especially when coupling effects at ferro- and antiferromagnetic interfaces occur, phenomena as exchange bias are found, first discovered by Meiklejohn and Bean more than 60 years ago \cite{meiklejohn_exchange_bias, nogues_exchange_bias, camley_exchange_bias}.

However, limitations are given by the resulting sample sizes. While magnetic effects can comfortably be studied and tuned by varying the thickness of films or size of spherical particles, it is getting more complex when sample sizes reach bulk dimensions. Hence, the processing of bulk materials with an internal nanostructure that still can be used for mass production is appreciated due to generally interesting coupling effects in nanocomposites consisting of two different magnetic materials \cite{bachmaier_tailoring}. To give an example, applying effects like the exchange bias in bulk samples will drastically improve the energy product, making this composites suitable for permanent magnet applications.
Covering the future demand for renewable energy requires an increasing amount of magnetic materials, where in the best case the use of rare earth elements should be avoided \cite{gutfleisch_magnetic_materials}. 

The investigation of nanostructured bulk materials can offer a huge amount of possibilities, but in particular the processing of these materials in bulk form is a challenging task. SPD by HPT is an opportunity to evade some processing limits. It allows the production of bulk alloys with non-equilibrium compositions below the melting point \cite{kormout}.
In this study the challenges of processing Fe-Cr alloys as a rare-earth-free alternative for permanent magnets are described. Different processing routes combining initial ball milling (BM) or arc melting (AM) prior to HPT deformation are investigated and an optimized processing route to obtain bulk, homogeneous nanocrystalline Fe-Cr alloys is found.

\section{Materials and Experimental}

Three different compositions consisting of nominal 30, 50 and 70 at.\% Fe ($ \pm$ 0.5 at.\%) were processed. For HPT and BM, conventional powders were used as starting materials, for AM experiments conventional flakes. 
All materials are stored and handled in Ar, BM was carried out in Ar-atmosphere. For BM, a planetary ball mill (Retsch  PM400) was used, with an 1:20 powder to ball ratio. Powders were milled at 400 rpm for a total milling time of 20 h.
A Buehler AM device under a Ti-gettered high purity Ar atmosphere was used to prepare homogeneously mixed ingots. The ingots were cut in discs with 8 mm diameter and 1 mm thickness, which are subsequently deformed by HPT.

The powders were hydrostatically compacted under Ar-atmosphere at a nominal pressure of 5 GPa and afterwards severely deformed at 7.5 GPa with 0.6 turns per minute, applying a high shear strain at room temperature (RT), 250°C or 400°C. The resulting bulk samples were 8 mm in diameter and had a thickness between 0.35 mm and 0.6 mm. The used setup is described in more detail elsewhere \cite{hohenwarter_HPT}.
For certain samples intermediate annealing between compaction and HPT deformation for 2 h at 1000°C under vacuum was also performed. Annealing of HPT deformed samples was performed for 1 h in air at different temperatures of 300°C, 400°C, 500°C, 600°C and 800°C.

Scanning electron microscopy (SEM) using back scattered electron diffraction (BSE), electron backscatter diffraction (EBSD) using transmission Kikuchi diffraction (TKD) and energy dispersive X-ray spectroscopy (EDX) were used for detailed microstructural characterization. 
If not specified otherwise, all microstructural investigations were performed at a radius r $\geq$ 2 mm to assure a homogeneously deformed microstructure \cite{martin_1}.
Vickers microhardness was measured along the diameter in steps of $ \Delta$r = 0.25 mm.
Synchrotron as well as X-ray diffraction measurements (XRD) were carried out in axial direction of the HPT disc using the facilities at Deutsches Elektronen-Synchrotron (DESY) and a Bruker Phaser D2 XRD equipment with Co-K$_{\alpha}$ radiation, respectively.

\section{Comparison of different processing routes}

\begin{figure}
  \includegraphics[width=\linewidth]{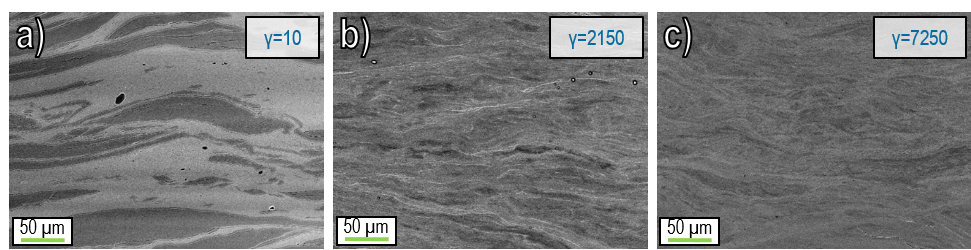}
  \caption{BSE images of Fe$_{50}$Cr$_{50}$ in tangential direction, deformed to different strains and temperatures: a) $ \gamma$= 10 and RT, b) $ \gamma$= 2150 and 400°C, c) $ \gamma$= 7250 and 250°C.}
  \label{fig:powder_HPT}
\end{figure}

\noindent \textbf{Processing of Fe-Cr powder blends by HPT deformation.} The phase diagram of Fe and Cr exhibits an immiscibility gap at low temperatures \cite{xiong_phase_2}. If compacting and subsequently deforming elemental powder blends with such a characteristic by HPT, the evolving microstructure can consist of a single phase supersaturated alloy or lead to a nanocomposite with partial supersaturation \cite{bachmaier_tailoring, kormout}. 
Powder blends of each Fe-Cr composition are HPT deformed and show similar characteristics. As an example, BSE images of the microstructures obtained after HPT deformation of the Fe$_{50}$Cr$_{50}$ powder blend are shown in Fig.\ref{fig:powder_HPT}. In Fig.\ref{fig:powder_HPT}a, an inhomogeneous microstructure with separated phases of Fe (bright) and Cr (dark) are visible at low strains after HPT deformation at RT. 
The limits of HPT processing are already reached due to an increase of hardness, leading to massive crack formation during HPT deformation. Elevated deformation temperatures usually lead to a final microstructure with larger grains, but allow the application of an higher amount of strain due to the reduced hardness of the processed material. Fig.\ref{fig:powder_HPT}b and c show the final microstructures of Fe$_{50}$Cr$_{50}$, deformed at 400°C and 250°C, respectively. Through the increased HPT deformation temperature, it was possible to apply a higher shear strain in both samples. Furthermore, co-deformation of Fe and Cr is visible and a significantly refined structure at higher shear strains is noticed. Final microhardness values at the outer edge of the HPT samples exceed 800 HV, which approaches the strength of the HPT anvils.

Successful HPT processing further depends on the composition of the Fe-Cr samples: samples with higher Cr content (Fe$_{30}$Cr$_{70}$) are observed to more likely fail during the deformation process due to shear band formation, a rapid increase in hardness or crack formation during processing. Even samples with lower Cr content (Fe$_{70}$Cr$_{30}$) exhibit crack formation during HPT deformation. 
To conclude, although co-deformation is observed, further refinement of the microstructure, which would be necessary to reach supersaturated states is impeded by the high hardness and crack formation of the Fe-Cr samples.

\begin{figure}[h]
  \includegraphics[width=\linewidth]{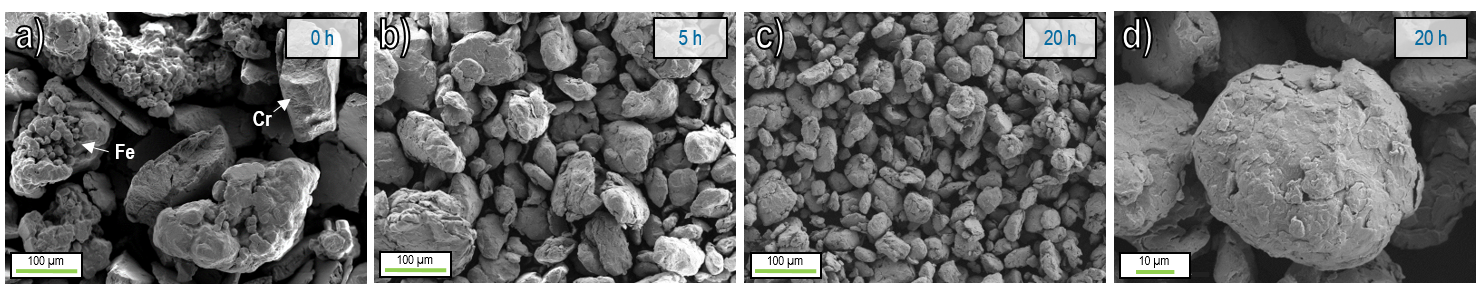}
  \caption{SEM images of BM Fe$_{50}$Cr$_{50}$-powders: a) 0 h, initial powders can be identified, b) after 5 h milling, c) after 20 h milling and d) magnified image after 20 h milling.}
  \label{fig:BM_evolution}
\end{figure}

\noindent \textbf{The effect of initial BM prior to HPT deformation.}
To circumvent the phase refinement step, BM is carried out to mechanically alloy the powders prior to HPT compaction and deformation. The milling process was interrupted several times to extract powders, which have been further investigated by XRD, SEM and EDX. In Fig.\ref{fig:BM_evolution}, SEM images of Fe$_{50}$Cr$_{50}$ powders after different milling times are shown. Without milling, separated Fe and Cr particles can be identified (Fig.\ref{fig:BM_evolution}a). With increasing milling time, the powder particles decrease in size and form a more globular shape. The particle size after 20 h of milling ranges between a few tens of $\mu$m to less than 100 $\mu$m.
EDX measurements show that the chemical composition converged to the original weight composition already after 15 h of milling time, indicating a homogeneous dispersion of Fe and Cr in the powder blends.
The same results are obtained for increased Fe or Cr contents (Fe$_{70}$Cr$_{30}$ or Fe$_{30}$Cr$_{70}$), indicating that alloying by BM was successful.

\begin{figure}
\floatbox[{\capbeside\thisfloatsetup{capbesideposition={right,center},capbesidewidth=6cm}}]{figure}[\FBwidth]
{\caption{XRD measurements of BM Fe$_{50}$Cr$_{50}$ powders with increasing milling times. Straight lines indicate the positions of theoretical diffraction maxima of Fe (green) and Cr (blue).}\label{fig:BM_XRD}}
{\includegraphics[width=0.55\textwidth]{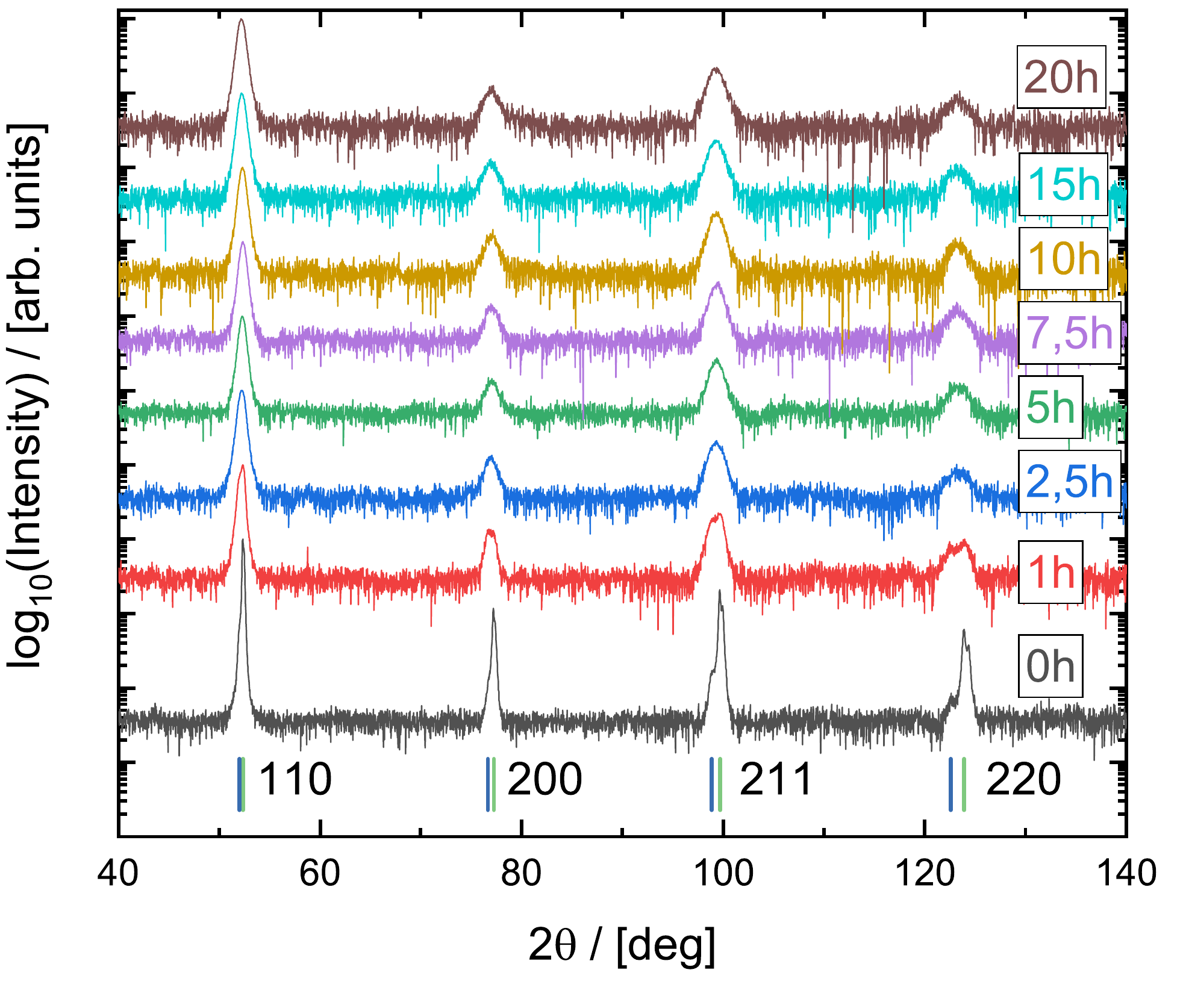}}
\end{figure}
XRD measurements of the BM Fe$_{50}$Cr$_{50}$ powder blends are shown in Fig.\ref{fig:BM_XRD}. Theoretical diffraction maxima of pure Fe (green, higher angle) and Cr (blue, lower angle) are calculated using the lattice constants taken from \cite{lattice_const}, which are indicated by straight lines and labeled by the miller indices. The resulting peak maxima after 20 h milling time stay between the calculated maxima of the pure elements. 
To prove the homogeneous distribution of Fe and Cr in the powders, the peak with highest intensity is analysed by Voigt functions. If the peak is fitted best by two functions, it can be assumed that Fe and Cr exists as two coexisting separated phases. However, the difference between experimental and fitted peak reaches a minimum, when plotting with just one single Voigt function. Therefore, it is assumed that a single phase Fe-Cr alloy powder is produced.
Further, the fitted peak positions are evaluated with respect to Vegard's law, revealing the same chemical composition as EDX measurements do. Indicating the consistency, the results are identical to the weight of the powders before starting the BM process. 

\begin{figure}[h]
  \includegraphics[width=\linewidth]{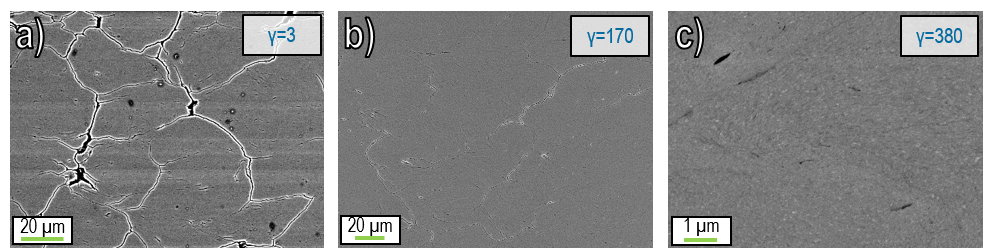}
  \caption{BSE images of compacted and HPT deformed Fe$_{50}$Cr$_{50}$ BM powder blends: a) $ \gamma$= 3 at RT, b) $ \gamma$= 170 at 250°C, c) $ \gamma$= 380 at 400°C.}
  \label{fig:BM_compaction}
\end{figure}

The BM powder blends are then compacted under Ar-atmosphere and deformed by HPT. In Fig.\ref{fig:BM_compaction}a, the microstructure of a compacted Fe$_{50}$Cr$_{50}$ sample after applying only a small amount of shear strain is depicted. The original particle surfaces can still be identified. Although the sample was HPT deformed to a maximum shear strain of $ \gamma$= 40 at RT, it was not enough to merge the particles on a microscale. By increasing the deformation temperature, it is possible to apply higher shear strains. Hence, former individual powder particles cannot be identified in Fig.\ref{fig:BM_compaction}b and \ref{fig:BM_compaction}c. A bulk material with homogeneous microstructure is reached for both deformation temperatures. 

Processing FeCr alloys by BM and subsequent HPT is also successfull for lower Cr contents (Fe$_{70}$Cr$_{30}$). Crack formation due to high hardness values can, however, not be prevented for samples containing 50 at.\% Cr and more.

\begin{figure}
  \includegraphics[width=0.49\linewidth]{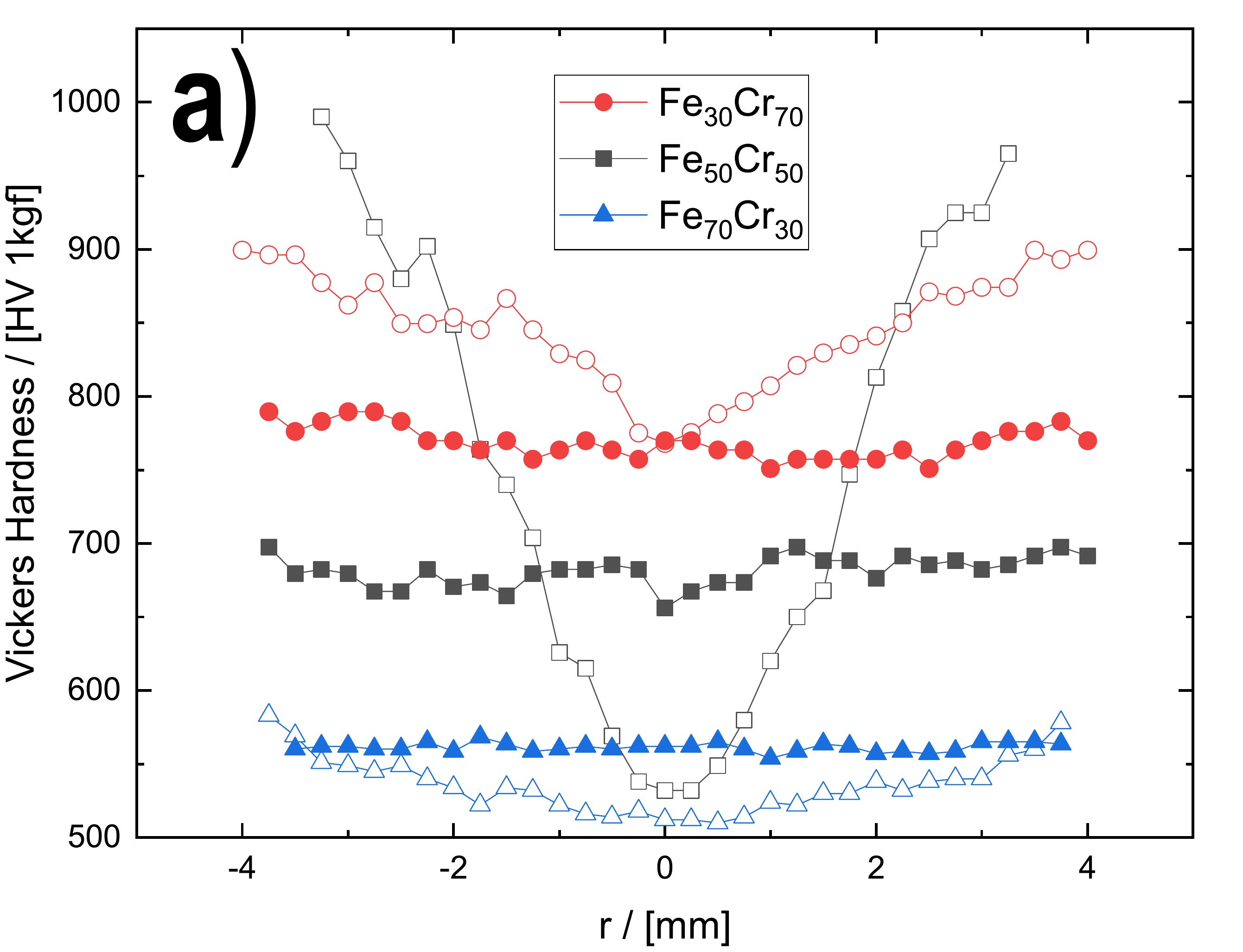}
  \includegraphics[width=0.49\linewidth]{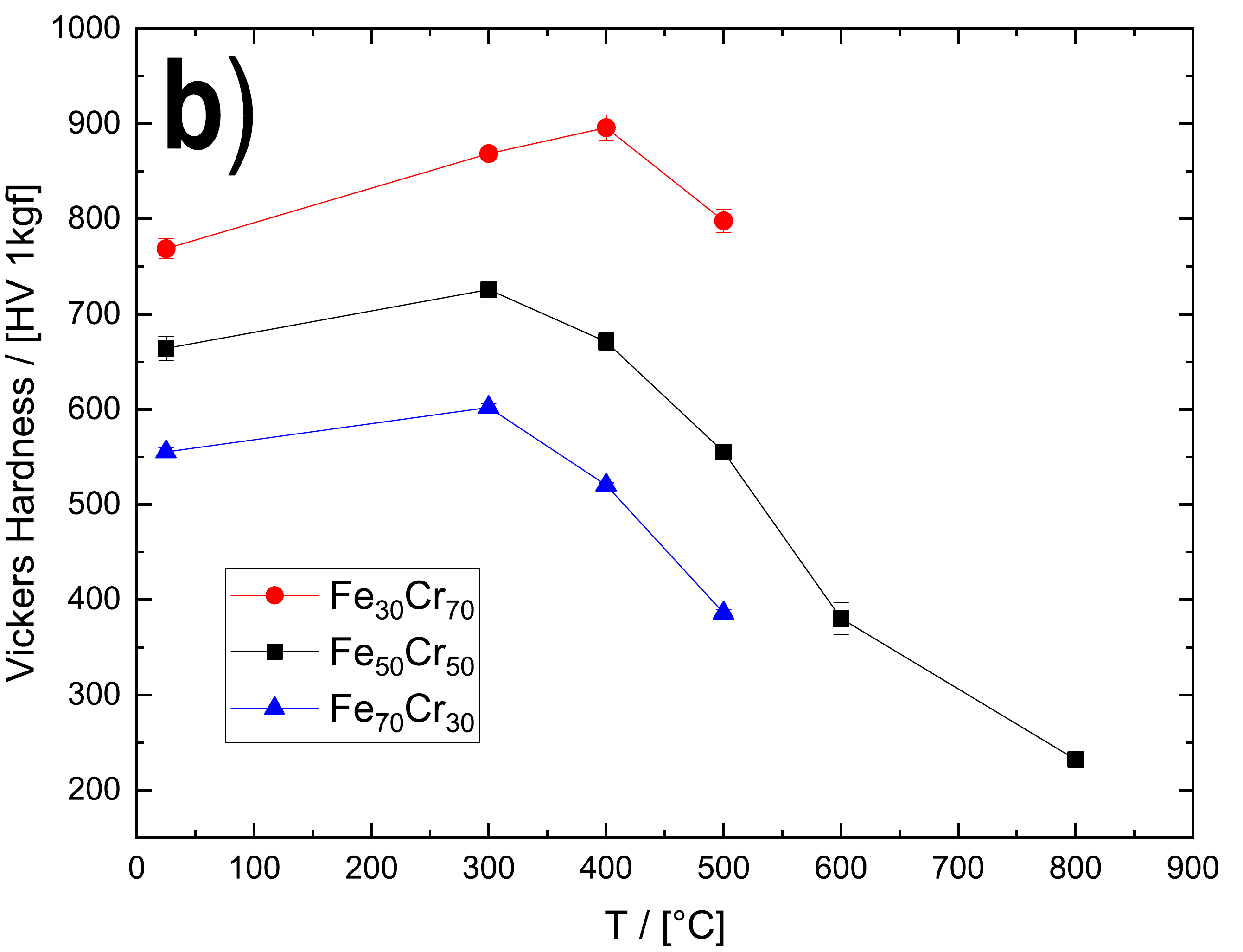}
  \caption{a) Microhardness of HPT deformed FeCr alloys: open symbols correspond to samples processed from BM powders and intermediate annealing before HPT deformation, filled symbols show HPT deformed AM samples. b) Microhardness of HPT deformed AM samples in as-deformed and annealed states.}
  \label{fig:hardness}
\end{figure}

Because BM powder blends already experience a high amount of deformation, annealing prior to HPT deformation is done to soften the powder particles. Homogeneous microstructures can be achieved for samples with a low Cr content (Fe$_{70}$Cr$_{30}$) after HPT deformation of these intermediate annealed samples. The microhardness after HPT deformation is less than 600 HV (Fig.\ref{fig:hardness}a). For higher Cr contents, a steadily increasing microhardness as a function of the radius with maximum values of 900 HV and 1000 HV can be found (Fig.\ref{fig:hardness}a), leading to massive crack formation during HPT deformation. 

 \begin{figure}
  \includegraphics[width=\linewidth]{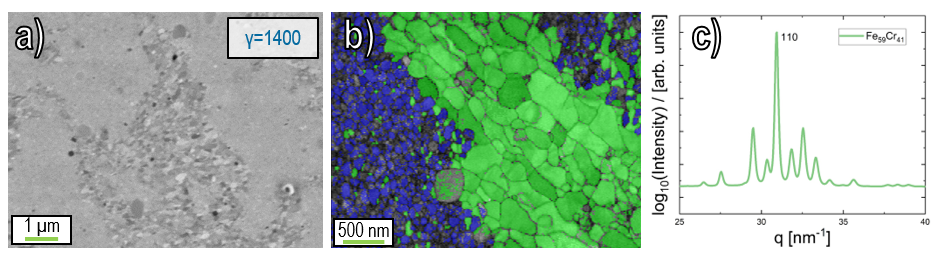}
  \caption{a) BSE image of HPT deformed Fe$_{59}$Cr$_{41}$ produced from BM powder blends with intermediate annealing prior to HPT deformation. b) TKD phase map with pattern quality overlay (blue: $ \sigma$-phase, green: bcc-\{Fe,Cr\}), c) Synchrotron XRD measurements indicate $ \sigma$-phase formation.}
  \label{fig:BM_VG_hpt}
\end{figure}

According to the phase diagram, a magnetic, but mechanically unwanted $ \sigma$-phase, can be found in a narrow region around 50 at.\% \cite{xiong_phase_2}. The formation of this phase by annealing should require an enormous time \cite{dubiel_sigma}. Fig.\ref{fig:BM_VG_hpt}a shows a BSE image of an HPT deformed Fe$_{59}$Cr$_{41}$ alloy with intermediate annealing prior to HPT deformation. A very fine structure as well as a coarse-grained phase can be seen. TKD measurements confirm the formation of $ \sigma$-phase after HPT deformation.
A phase map of a Fe$_{59}$Cr$_{41}$ alloy after HPT deformation is shown in Fig.\ref{fig:BM_VG_hpt}b.
The phase map is overlapped with the pattern quality image to illustrate grain boundaries. The grain size of the Fe phase is significantly larger compared to the $ \sigma$-phase.
Fig.\ref{fig:BM_VG_hpt}c shows the results from synchrotron XRD measurements. Beside the main bcc 110 reflection, several peaks are observed. They can be associated to the formation of the $ \sigma$-phase.
In literature, it is shown that plastic deformation favours the formation of the $ \sigma$-phase \cite{dubiel_sigma}, which might be the reason for the quick $ \sigma$-phase formation during HPT deformation.\\

\begin{figure}[h]
  \includegraphics[width=\linewidth]{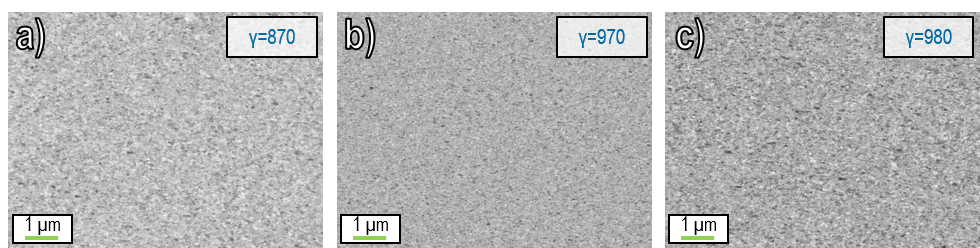}
  \caption{BSE images of HPT deformed AM FeCr alloys. a) Fe$_{30}$Cr$_{70}$, b) Fe$_{50}$Cr$_{50}$, c) Fe$_{70}$Cr$_{30}$.}
  \label{fig:AM_hpt}
\end{figure}

\noindent \textbf{HPT deformation of AM alloys.}
The Fe-Cr alloys produced by AM were HPT deformed at RT. It was possible to produce samples with a homogeneous microstructure for all chemical compositions. BSE images of the final microstructures are shown in Fig.\ref{fig:AM_hpt}. The corresponding microhardness along the radius is displayed in Fig.\ref{fig:hardness}a with filled symbols for the HPT deformed AM alloys. When comparing to the intermediate annealed BM powder samples, major differences are found. For the HPT deformed BM powder samples one can see a hardness increase with the radius, indicating that a steady state microstructure in the alloys (e.g. completely finished intermixing, steady state grain size in phases) is not reached. Further, the $ \sigma$-phase formation leads to a massive hardness increase for the Fe$_{50}$Cr$_{50}$ sample. On the contrary, the HPT deformed AM alloys display a constant microhardness over the sample radius, hence a homogeneous microstructure in the steady state is reached.

To produce a hard magnetic material with a high energy product induced by exchange coupling, a structure with nanometer-sized Fe and Cr must be achieved. 
Therefore, after successful processing, annealing experiments are performed to decompose the microstructure and form a nanocomposite.
BSE images after annealing (not shown), display coarsening of the structure, corresponding hardness measurements are shown in Fig.\ref{fig:hardness}b. For all three compositions, an increase in hardness is found for low annealing temperatures, which might be related to decomposition and/or an effect named hardening by annealing \cite{renk}. 
When the annealing temperatures are further increased, the microhardness decreases. The microhardness as well as the hardening effect is further shifted to higher annealing temperatures with increasing Cr content. Detailed characterization of the annealed microstructures and resulting magnetic properties will be carried out in a further study. 

\section{Conclusion}

The challenges of processing FeCr alloys by HPT are presented using different processing routes. It is shown that HPT processing of conventional powder blends leads to a microstructure of very fine but separated Fe and Cr phases. Crack formation is a limiting problem. Using BM, Fe-Cr alloys for 
three different compositions are successfully manufactured. Subsequent HPT deformation is performed after an intermediate annealing step. However, an unwanted and unusual fast growing $ \sigma$-phase is identified after HPT deformation, which limits the formation of an homogeneous microstructure by an immense increase in hardness.
Finally, Fe$_{30}$Cr$_{70}$, Fe$_{50}$Cr$_{50}$ and Fe$_{70}$Cr$_{30}$ samples with an homogeneous nanocrystalline microstructure are obtained using AM prior to HPT deformation. Furthermore, annealing of as-deformed samples, leads to a hardening of the microstructure which is found for all chemical compositions.

\section{Acknowledgements}

The synchrotron measurements leading to these results have been performed at PETRA III: P07 at DESY, a member of the Helmholtz Association. The authors gratefully acknowledge the assistance by N. Schell and for the help with AM, BM and XRD measurements S. Ketov, N. Chawake and J. Todt, respectively.
This project has received funding from the European Research Council (ERC) under the European Union’s Horizon 2020 research and innovation programme (Grant No. 757333).

\end{document}